\documentclass[%
reprint,
nofootinbib,
amsmath,amssymb,
aps,
]{revtex4-2}

\usepackage{graphicx} % Include figure files
\usepackage{dcolumn} % Align table columns on decimal point
\usepackage{bm} % bold math
\usepackage{hyperref} % add hypertext capabilities
\usepackage{color}
\usepackage{soul}
\usepackage[utf8]{inputenc}
\usepackage{enumitem}
\usepackage{dsfont}
\usepackage[dvipsnames]{xcolor}
\usepackage{physics}
\usepackage{float}
\usepackage{makecell}
\usepackage{upgreek}
\usepackage{lipsum, babel}
\usepackage{indentfirst}
\usepackage{fdsymbol}
\usepackage{ulem} 
\graphicspath{ {./} }

\definecolor{bittersweet}{rgb}{1.0, 0.67, 0.1}

\begin{document}

\title{Feedback Enhanced Phonon Lasing of a Microwave Frequency Resonator}

\author{Peyman Parsa, Prasoon Kumar Shandilya, David P.\ Lake, Matthew E.\ Mitchell, and Paul E.\ Barclay}

\email[Paul~E.\ Barclay: ]{Corresponding author pbarclay@ucalgary.ca}
\affiliation{
Institute for Quantum Science and Technology, University of Calgary, Calgary, Alberta T2N 1N4, Canada}

\date{\today}

\begin{abstract}
The amplitude of self-oscillating mechanical resonators in cavity optomechanical systems is typically limited by nonlinearities arising from the cavity's finite optical bandwidth. We propose and demonstrate a feedback technique for increasing this limit. By modulating the cavity input field with a signal derived from its output intensity, we increase the amplitude of a self-oscillating GHz frequency mechanical resonator by $22\%$ (increase in coherent phonon number of $50\%$) limited only by the achievable optomechanical cooperativity of the system.
%We demonstrate enhanced amplification in optomechanical systems using RF feedback. We use the RF signal detected on a photoreceiver. The signal is amplified, phase shifted, and fed back into a phase modulator. We show that this method could modify the amplitude of mechanical motions obtained from phonon lasing. We used a diamond microdisk that can support mechanical modes up to GHz range to implement this method. 
This technique will advance applications dependent on high dynamic mechanical stress, such as coherent spin-phonon coupling, as well as implementations of sensors based on self-oscillating resonators.
%Having control over the amplitude and the frequency of the oscillations.
%The thermal amplitude is $(GA/wm)^2$ = 8.64e-7. 
\end{abstract}

\maketitle

\section{Introduction}

Feedback is a key element of many classical and quantum technologies ranging from control of quantum systems \cite{Vijay2012}, implementations of mechanical sensors such as atomic force microscopes \cite{Albrecht1991}, and highly coherent optical sources  \cite{Dahmani1987, Jin2021}. Cavity optomechanical systems \cite{RevModPhys.86.1391} commonly use various forms of feedback \cite{Guo2019, Wilson2015, Rossi2018, Ernzer2023} to damp mechanical motion of resonators interacting with an optical cavity and cool them to their quantum ground state. Feedback can also be used to amplify mechanical motion in these systems to probe the limits of nanomechanical sensors \cite{Buks2006, Guha2020, Manzaneque2023, Liu2021, Bemani2022}, generate frequency combs \cite{PhysRevLett.127.134301} and rf and microwave oscillators \cite{HosseinZadeh2010, HosseinZadeh2008, Jiang2012, Han2014, Huang2017, Mercade2020}, and study nonlinear dynamics \cite{Krause2012}. Self-oscillating phonon lasers have also  played a role in optomechanically driving electronic spin qubits associated with defects in materials such as diamond \cite{shandilya2021optomechanical}. 

%Advances in quantum technology are reliant upon quantum control, with applications such as quantum networking \cite{Kimble2008}, sensing \cite{RevModPhys.89.035002}, computing \cite{Ladd2010} and communication \cite{RevModPhys.74.145} relying upon it. Colour centers in solids have been proven to be useful components of quantum technologies \cite{Awschalom2018}. Nitrogen vacancy (NV) centers in diamond are one of these promising colour centers that could provide a means of spin-mechanics coupling \cite{Shandilya2021}. Spin-mechanics coupling is typically weak and this hinders the spin-optomechanical control of spins.

By localizing optical fields near or within a mechanical resonator, cavity optomechanical systems provide coupling between optical and mechanical degrees of freedom that is essential for using feedback to control resonator dynamics.  The presence of a delay in the system's optical response to resonator displacement--dynamical optomechanical backaction--allows the optomechanical coupling to modify the resonator's effective mechanical dissipation \cite{aspelmeyer2014cavity}. Both internal feedback from the cavity mode's delayed optical response to changes in resonator geometry \cite{Park2009, Schliesser2009, ref:chan2011lcn}, and delayed external feedback derived from the measured \cite{PhysRevLett.83.3174, PhysRevLett.80.688} or re-circulated \cite{Ernzer2023} cavity field have been used to cool mechanical resonators. Similarly, excitation of coherent resonator motion through these forms of feedback  has been demonstrated \cite{HosseinZadeh2010, PhysRevApplied.17.034020, PhysRevLett.83.3174}.  Cavity-less optomechanical systems can also harness feedback: for example, adaptive control has been employed for cooling levitated mechanical objects \cite{PhysRevLett.122.223602}, and photothermal optmoechanical coupling allows modification of resonator dynamics by microscope field gradients \cite{Jayakumar2021} without the aid of a cavity.   

Although phonon lasing typically increases the resonator amplitude by several orders of magnitude, its maximum displacement is clamped by nonlinear optomechanical effects arising when the mechanically induced change in cavity frequency exceeds the cavity linewidth \cite{PhysRevA.86.053826}, limiting its potential for many of the applications discussed above. Enhancing the dynamics of self-oscillating resonators by modulating the continuous wave field that parametrically excites their mechanical motion has been demonstrated in studies of frequency injection locking \cite{injection-hosseinzadeh}. Here we examine whether modulation of the drive field allows the self-oscillation amplitude to be increased. We show that the amplitude of a self-oscillating optomechanical resonator can be enhanced with the aid of linear external feedback that complements the cavity's internal backaction and compensates for the optomechanical nonlinearities limiting the oscillation amplitude. This external feedback is similar to linear velocity feedback used in Ref.\ \cite{PhysRevLett.122.223602} for nanoparticle levitation, and is applied here to control the self-oscillations of a GHz frequency diamond microdisk cavity optomechanical device. After being driven into phonon lasing, the measured mechanical motion of the resonator is amplified, phase shifted, and input into a phase modulator that modulates the input laser frequency so that it follows the optical cavity frequency set by the mechanical displacement. This process effectively reduces dynamical backaction and enables the self-oscillating resonator to accumulate a higher number of phonons for a given input power.

\begin{figure}
    \includegraphics[width = \linewidth]{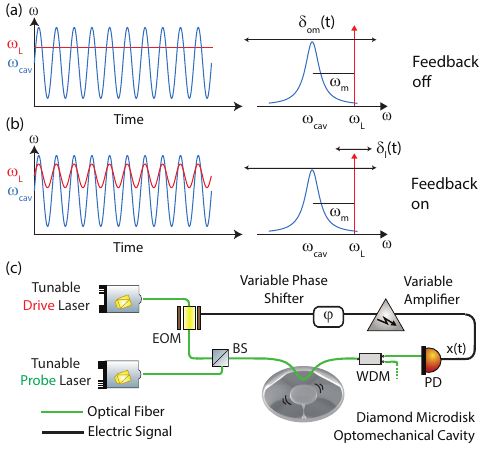}
    \caption{(a,b) Temporal evolution of the optical cavity (blue) and laser (red) frequencies for an oscillating cavity optomechanical system, with and without external feedback used to shift the laser frequency so that it follows the instantaneous optomechanical frequency shift. (c) Experimental system used to implement the laser frequency feedback scheme.   }
    \label{fig:fig1}
\end{figure}

\section{Theory of feedback enhanced of parametric optomechanical driving}
\label{sec:theory}

Modification of mechanical resonator dynamics through coupling to an optical cavity driven by a continuous wave laser arises when the displacement of the resonator by the cavity field modifies the cavity mode frequency, which in turn affects the field built up in the cavity. If the delay set by the cavity bandwidth is comparable to or longer than the mechanical oscillation period, this optomechanical backaction results in damping ($\Gamma_\text{om} > 0$) or anti-damping ($\Gamma_\text{om} < 0$) of the resonator motion depending on whether the laser is red ($\Delta < 0$) or blue ($\Delta > 0$) detuned from the cavity mode. The mechanical resonator is driven into self-sustained oscillation when the energy transfer rate $\Gamma_\text{om}$ between the cavity field and the resonator is equal to the resonator's internal dissipation rate $\Gamma_\text{m}$ \cite{PhysRevLett.96.103901}, and in a linearized theory of optomechanics, modifies the resonator's phonon number from its thermal occupation $n_\text{th}$ to $n = n_\text{th} \Gamma_\text{m}/(\Gamma_\text{m} + \Gamma_\text{om})$. This expression reveals the divergence of $n$ and onset of self-oscillation when $\Gamma_\text{om} = -\Gamma_\text{m}$, a regime  that is most efficiently accessed when $\Delta = \omega_\text{m}$ so that the parametric process of scattering a laser photon into the lower frequency cavity mode while generating a phonon is resonant.

In practice, the amplitude of a self-oscillating resonator is clamped when the instantaneous optomechanical frequency shift $\delta_\text{om}(t) = G x(t)$ of the cavity mode frequency induced by the oscillations is comparable to the cavity linewidth, as determined by the optomechanical coupling coefficient $G = -d\omega_\text{o}/dx$ and the oscillation amplitude $A$. This is illustrated in Fig.\ \ref{fig:fig1}(a), which shows how a laser nominally blue detuned $\Delta = \omega_\text{m}$ to maximize anti-damping is shifted away from this detuning by $\delta_\text{om}(t)$ when $A$ is large. Previous numerical studies have explored how varying the cavity optomechanical system's parameters affect $A$ \cite{ref:poot2012bls}. Here we explore an alternative concept: whether $A$ can be enhanced using feedback to  dynamically compensate for $\delta_\text{om}(t)$ by modulating the input laser frequency. 

This approach is illustrated in Fig.\ \ref{fig:fig1}(b): given knowledge of $x(t)$, the laser frequency $\omega_\text{l}$ is constantly adjusted by $\delta_\text{l}(t)$ to reduce its deviation away from $\Delta = \omega_\text{m}$.  Our experimental implementation of this scheme is shown in Fig.\ \ref{fig:fig1}(c). A weak laser input to an optical cavity through an optical waveguide provides readout of $x(t)$ via photodetection of its transmission.  This electronic signal is fed-back to an electro-optic modulator that dynamically shifts the phase of a strong field $\alpha$ input to a second cavity mode that parameterically drives the cavity's mechanical resonance into self-oscillation for sufficient optical power $|\alpha|^2$ and appropriate detuning $\Delta$. In the experiment describe below, a diamond microdisk whispering gallery mode cavity evanescently coupled to a fiber taper waveguide is used. This system is described by equations of motion,
\begin{align}
        &\dot{\alpha} = -\frac{\kappa}{2} \alpha + i(\Delta + Gx)\alpha + \sqrt{\kappa_\text{ex}}\alpha_\text{in} e^{+ ifGx(t-\tau)/\omega_\text{m}}, \\
        &\ddot{x} + \Gamma_\text{m}\dot{x} + \omega_\text{m}^2 x = \frac{\hbar G}{m_\text{eff}}|\alpha|^2, 
    \label{eq:eqappa2}
\end{align}
with feedback captured by the time varying shift $f G x(t-\tau)/\omega_\text{m}$ in the phase of the input field $\alpha$, where $\tau$ is the delay of an electronic feedback line and $f$ is the feedback strength, which are controlled by a variable phase shifter and a variable gain amplifier, respectively. Here $\kappa_\text{ex}$ is the coupling rate between the cavity and the input waveguide, and $m_\text{eff}$ is the resonator mode's effective mass.

The nonlinear dynamics of this system can be analyzed by generalizing the theory of Marquardt et al.\  \cite{PhysRevLett.96.103901} to include external feedback. In steady state, the power scattered into the resonator from optical radiation pressure is equal to the resonator's dissipated power, so that $\left\langle\ddot{x}\dot{x}\right\rangle = 0$. This leads to, 
\begin{align}
\label{eq:eq1}
\frac{2\omega_\text{m}}{\kappa C_\text{om}} &\approx \frac{1}{|z_0|^2}\sum_{j=-\infty}^{\infty}\frac{J_j(\beta)J_{j+1}(\beta)}{\beta/2}\mathfrak{Im}\{z_j z_{j+1}^*(1-fe^{i\varphi})\},
%\frac{\gamma_\text{m}}{\omega_\text{m}} &\approx \frac{g_0^2}{\omega_\text{m}^2}\sum_{j=-\infty}^{\infty}\frac{J_j(\beta)J_{j+1}(\beta)}{\beta}\mathfrak{Im}\{z_j z_{j+1}^*(1-fe^{i\varphi})\},
% n_\text{cav} &\approx \sum_{j = -\infty}^{\infty} J_j(\beta)^2z_j^*z_j,
\end{align}
where $\beta = {GA \sqrt{1+f^2-2f \cos(\varphi)}}/{\omega_\text{m}}$ is the effective optomechanical modulation index,  $z_j = {\sqrt{\kappa_\text{ex}}\alpha_\text{in}}/[{{\kappa}/{2}-i(\Delta+j\omega_\text{m})}]$ is the cavity mode response at sideband frequency $\omega_\text{l} + j \omega_\text{m}$, $|z_0|^2$ is the intracavity photon number before the onset of self-oscillations, and $C_\text{om} = 4g_0^2|z_0|^2/\kappa\Gamma_\text{m}$ is the optomechanical cooperativity before the onset of self-oscillations.  Physically, Eq.\ \eqref{eq:eq1} compares $\Gamma_\text{m}$ to the sum of the phonon generation rates associated with optomechanical scattering between the $j+1$ and $j$ optical sidebands. 

\begin{figure}
  \includegraphics[width = \linewidth]{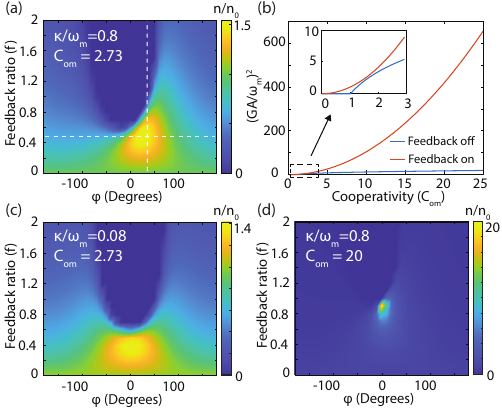}
  \caption{(a) Theoretically predicted phonon occupation of the mechanical resonance as a function of feedback parameters, normalized by the number of phonons during pure self-oscillation ($f = 0$) for the device parameters used in this work ($\kappa/\omega_\text{m} = 0.8$) and an optimally blue detuned drive field with intensity corresponding to $C_\text{om} = 2.73$. Dashed lines correspond to parameter ranges studied experimentally below. (b) Predicted normalized phonon occupation for varying $C_\text{om}$ with optimized feedback parameters (red) and feedback turned off (blue). (c) and (d) repeat part (a) for $\kappa/\omega_\text{m} = 0.08$ and  $C_\text{om} = 2.73$, and $\kappa/\omega_\text{m} = 0.8$ and $C_\text{om} = 20$, respectively.}
  \label{fig:fig2}
\end{figure}

Feedback manifests in the term $(1-fe^{i\varphi})$ in Eq.\ \eqref{eq:eq1} and a modification of $\beta$,  affecting the optomechanical scattering rates between sidebands. As written, Eq.\ \eqref{eq:eq1} can be conveniently solved numerically for mechanical amplitude $A$ as a function of $C_\text{om}$, allowing the influence of $f$ and $\varphi$ to be studied for a given $\kappa/\omega_\text{m}$.  Figure \ref{fig:fig2}(a) shows the result of this calculation for $\kappa/\omega_\text{m} = 0.8$ and $C_\text{om} = 2.73$, chosen since they correspond to operating conditions of the cavity optomechanical system studied experimentally below. We see that the oscillation amplitude, expressed in terms the phonon occupation of the mechanical resonator mode $n \propto (GA/\omega_\text{m})^2$, possesses a maximum that is $\sim 1.5\, \times$ larger than the amplitude $n_0$ in absence of feedback ($f = 0$).  The dependence of $n/n_0$ on $f$ and $\varphi$ is consistent with the behaviour of optomechanical damping: dynamical back-action is responsible for phonon lasing \cite{RevModPhys.86.1391}, but also limits the oscillation amplitude \cite{PhysRevA.86.053826}.  It is worth noting that the points where the ratio is zero indicate that the system is not self-oscillating and the number of coherent phonons is zero.
Note that for all of these calculations $\Delta = \omega_\text{m}$, and that the input field strength $\alpha$ and coupling rate $\kappa$ do not need to be specified in this analysis.

To further explore the feedback's behavior, we solve Eq.\ \ref{eq:eq1} for varying cooperativity, optimizing the feedback parameters at each $C_\text{om}$ to maximize $n$. Figure \ref{fig:fig2}(b) shows the result. This clearly indicates that in absence of feedback ($f = 0$),  as $C_\text{om}$ increases, which for a given set of cavity parameters corresponds to increasing the number of intracavity drive photons $n_\text{cav}$, the resonator amplitude saturates. Conversely, when feedback is used, $n$ grows as $C_\text{om}$ increases. One can show that at the optimal point, $A$ is linearly proportional to the cooperativity $GA/\omega_\text{m} \propto C_\text{om}$ with the proportionality constant to be maximized over $f$ and $\varphi$. This is seen in Fig.\ \ref{fig:fig2}(b) where $\left(GA/\omega_\text{m}\right)^2$ scales quadratically with $C_\text{om}$.  This shows that adding feedback allows the phonon number to be enhanced by orders of magnitude, limited in practice by the experimentally achievable $C_\text{om}$. We also note that self-oscillation is possible using feedback for $C_\text{om} < 1$, as shown in the inset to Fig.\ \ref{fig:fig2}(b).  {Figures \ref{fig:fig2}(c) and \ref{fig:fig2}(d) illustrate the impact of device parameters on the feedback enhancement. For $\kappa/\omega_\text{m} = 0.08$ (Fig.\ \ref{fig:fig2}(c)) corresponding to increasing $Q_\text{o}$ or $\omega_\text{m}$ by an order of magnitude, the maximum phonon occupation is not significantly affected if $C_\text{om}$ is kept constant. However, note that higher $Q_o$ allows achieving higher $C_\text{om}$ for a given $n_\text{cav}$. When $C_\text{om}$ is increased to 20 while $\kappa/\omega_\text{m}$ remains constant (Fig.\ \ref{fig:fig2}(d)), the maximum phonon occupation increased by nearly an order of magnitude. } 

\section{Experimental Demonstration}
\label{sec:experimental setup}
% \begin{figure*}
%   \centering
%   \includegraphics[width = \linewidth]{Figures/Figure 3.pdf}
%   \caption{(a) Input fields in frequency domain. $\Delta$: Probe Detuning (b) Schematic of the experimental setup. PD: Photoreceiver, EOM: Electro-optic Phase Modulator, BS: Beam-splitter, WDM: Wavelength-division Multiplexer \pb{Combine elements of this with Fig. 1.}}
%   \label{fig:fig3}
% \end{figure*}
We next experimentally demonstrate feedback enhanced optomechanical self-oscillation. The cavity optomechanical system used here consisted of a diamond microdisk previously used for coherent photon--phonon coupling \cite{lake2021processing, lake2020two} and spin--optomechanics \cite{shandilya2021optomechanical}. This device supports a $\omega_\text{m}/2\pi = 2.1\,\text{GHz}$ radial breathing mode mechanical resonance that interacts strongly with optical whispering gallery modes of the device. A fiber taper waveguide evanescently couples to optical modes used for driving and probing the mechanical motion of the microdisk. The drive mode (wavelength $\lambda_\text{d} = 1563.4\, \text{nm}$) has quality factor  $Q_\text{o,d}  = 1.1\times10^5$ sufficiently high to place it near the sideband resolved regime, while the probe mode ($\lambda_\text{p} = 1509.5\,\text{nm}$) has lower $Q_\text{o,p} = 1.1 \times 10^4$, allowing its field to instantaneously respond to the motion of the mechanical resonator. Coherent photon--phonon coupling between the drive mode and the radial breathing mode can achieved thanks to their high optomechanical coupling rate $g_0/2\pi = 25$ kHz, the mechanical resonance's high quality factor $Q_\text{m} = \omega_\text{m}/\Gamma_\text{m} =  4,300$, and diamond's ability to support high intensity fields before the cavity exhibits heating instability and nonlinear absorption. In the measurements presented below, a drive field of approximately $n_\text{cav} = 0.9 \times 10^6$ photons (1.4 mW dropped power) is used to realize photon-enhanced $C_\text{om} \approx 2.73$. In all measurements, the probe laser power is sufficiently low for it to have no effect on the mechanical resonance dynamics.

To implement the feedback scheme, the RF component of the photodetected probe field signal is amplified, delayed, and input to an electro-optic modulator (EOM) that shifts the phase of the drive field. At the fiber taper input the probe and drive fields are combined using a 90:10 fiber coupler, and at the  output they are separated using a wavelength division multiplexer (Montclair Fiber MFT-MC-51-30-AFC/AFC-1). The transmitted probe field is monitored using a fast photoreceiver (Thorlabs RXM25AA) whose output is filtered near $\omega_\text{m}$ (passband 2.0-2.3 GHz, Mini Circuits VBFZ-2130-S+) and measured using an RF power detector (Mini Circuits ZX47-50LN-S+) to generate the signal input to the feedback circuit. Two amplifiers (Mini Circuits ZKL-33ULN-S+ \& ZX60-83LN-S+) and a variable attenuator (Mini Circuits ZX73-2500-S+) realize a variable gain amplifier that varies $f$ and a phase shifter (Mini Circuits JSPHS-2484+) that varies $\tau$. The resulting signal is used to drive the EOM (EOSPACE PM-5S5-20-PFA-PFA-UV-UL).

\begin{figure}
    \centering
    \includegraphics[width = \linewidth]{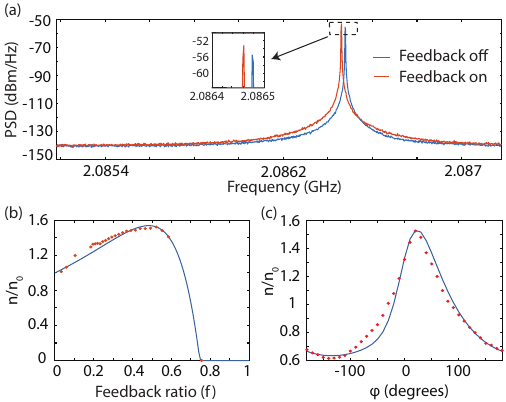}
    \caption{Experimental demonstration of feedback enhanced optomechanical self-oscillation. (a) Power spectral density of the photodetected signal showing the self-oscillation of the microdisk radial breathing mode with and without electronic feedback turned on. The signal is measured from the probe laser transmission through the fiber taper waveguide when it is tuned near resonance with the probe cavity mode, and the drive laser is detuned by $\-\omega_\text{m}$ from the drive mode. In (b) and (c) the dependence of the area under the self-oscillation resonance is plotted for varying $f$ and $\varphi$, respectively, and compared with theoretical predictions. In each plot the non-varying feedback parameter is fixed at its optimal value.}    \label{fig:fig4}
\end{figure}

Figure \ref{fig:fig4} shows the measured power spectral density with and without feedback when the device is excited into self-oscillation by the blue detuned drive laser.  The area under the peak is a measure of the mechanical power and corresponding phonon occupation and oscillation amplitude. An increase in mechanical power corresponding to  $n/n_0 \approx 1.5$ is obtained when feedback is on and the feedback parameters are optimized. Figures \ref{fig:fig4}(b) and \ref{fig:fig4}(c) show the dependence of $n/n_0$ on these parameters, revealing a clear maxima at $f \approx 0.5$ and $\varphi \approx 20^\circ$. A cooperativity of $C_\text{om} \approx 2.73$ is estimated for these measurements by referencing the drive field input power to its value at the onset of self-oscillations (corresponding to $C_\text{om} = 1$). 

Also shown in Figs.\ \ref{fig:fig4}(b) and \ref{fig:fig4}(c) is the theoretically predicted dependence of $n/n_0$ on the feedback parameters. Each of these plots are slices through the parameter space shown in Fig.\ \ref{fig:fig2}(a), with the non-varying feedback parameter fixed at its optimal setting. When gain is varied in Fig.\ \ref{fig:fig1}(b), 
%with the phase shift optimized ($\varphi \approx 20^\circ$), 
we see that, as discussed in Sec.\ \ref{sec:theory}, the number of phonons initially increases with increasing $f$, until reaching a maximum at $f = 0.5$. For larger $f$ the phonon number decreases, and for $f > 0.75$ we find that the number of coherently generated phonons becomes zero, meaning that there is no self-oscillation and that the effective cooperativity becomes less than one. When the feedback's phase shift is varied, as shown in Fig.\ \ref{fig:fig1}(c), we see that the enhancement decreases gradually on either side of its maxima.  Comparing both sets of measurements with the theoretically predicted values of $n/n_0$ we find excellent agreement, including the observed asymmetry of ${n}/{n_0}$ with respect to $\varphi$, which is found to be due to the non-zero cavity linewidth $\kappa$.

\section{Discussion and Conclusion}\label{sec:conclusion}

The enhancement in phonon occupation demonstrated here is not limited by fundamental mechanisms. Rather, as discussed above, $n$ is predicted to increases quadratically with $C_\text{om}$, in contrast to its behaviour in conventional phonon lasing where $n$ saturates due to dynamical back-action. Increasing $C_\text{om}$ beyond the value demonstrated here can be achieved most directly through operating at higher laser input power. However, in practice this is limited by residual heating of the diamond device that can lead to thermal instability of the optical mode \cite{lake2018optomechanically}. Alternatively, $C_\text{om}$ can be increased by improving the device parameters. Reducing its mechanical dissipation is particularly desirable since the scattering processes governing the dependence of $n$ on $C_\text{om}$ in Eq.\ \eqref{eq:eq1} are not affected by $\Gamma_\text{m}$. For example, increasing $Q_\text{m}$ to 9,000 observed in Ref.\ \cite{mitchell2016single} would immediately offer a further $2\times$ increase in phonon occupation. Implementing the feedback scheme using optomechanical crystals \cite{ref:eichenfield2009oc}, which have been demonstrated with $Q_\text{m} > 10^{10}$ \cite{MacCabe2020} would offer orders of magnitude higher enhancement.

The increase in mechanical self-oscillation amplitude accessible using this feedback scheme will enable more efficient optomechanical driving \cite{shandilya2021optomechanical} of electronic spin systems such as diamond nitrogen vacancy (NV) and silicon vacancy (SiV) color centers, enabling spin--optomechanical control to enter the regime of coherent spin-phonon coupling \cite{MacQuarrie2015}. Enhanced self-oscillation amplitude will also provide access to rich nonlinear optomechanical dynamics \cite{Krause2012} and assist in observing nonlinear nanomechanical effects \cite{WestwoodBachman2022} of interest for applications such as generating squeezed states \cite{Lue2015}. Implementing this scheme using high signal to noise homodyne detection may allow excitation of thermal states into self-oscillation.

%Feedback can also be used for cooling the mechanical resonators, as shown in \cite{PhysRevLett.80.688} using MHz frequency devices, provided that the signal to noise of the measurement of the resonator's thermal Brownian motion is sufficiently high. Although this is technically challenging at GHz frequencies using direct photodetection, it may be possible using homodyne techniques Ref.\ \cite{XX}. 

\begin{acknowledgments}
We wish to acknowledge support from Alberta Innovates (Strategic Research Project), the Government of Alberta Major Innovation Fund, the Canadian Foundation for Innovation, the National Research Council (NanoInitiative Program), and the Natural Sciences and Engineering Research Council of Canada (Discovery Grant and Strategic Partnership Grant programs).
\end{acknowledgments}

\bibliography{nano_bib, quantnanobib}

\end{document}